# PFA-NS: Power-Fading-Aware Noise Shaping Enabled C-Band IMDD System with Low Resolution DAC

Xiaobo Zeng[1, *], Liangcai Chen[1], Pan Liu[1], Ruonan Deng[2]

[1]*Hunan Province Key Laboratory of Credible Intelligent Navigation and Positioning, Key Laboratory of Intelligent, Computing and Information Processing of Ministry of Education, National Center for Applied Mathematics in Hunan, Xiangtan University, Xiangtan, 411105, China.*
[2]*College of Meteorology and Oceanology, National University of Defense Technology, Changsha, 410073, China.*
*\*Corresponding Author: xiaobo.zeng@xtu.edu.cn*

**Abstract:** We propose and demonstrate a power-fading-aware noise-shaping technique for C-band IMDD system with low resolution DAC, which shapes and concentrates quantization noise within the fading-induced notch areas, yielding 94% improvement in data-rate over traditional counterpart. © 2026 The Author(s)

## 1. Introduction

Driven by the proliferation of bandwidth-intensive network applications, global internet traffic is experiencing rapid growth, significantly heightening the demand for high-capacity photonic data-center interconnections (DCIs)[1, 2]. Owing to the stringent cost and power constraints in DCI deployments, intensity modulation and direct detection (IMDD), compared with the coherent alternatives, offers distinct advantages in simplicity and cost-effectiveness, positioning it as the dominant solution for data center environments[3, 4]. For standard single mode fiber (SSMF), the attenuation in C-band exhibits significantly lower than that in the O-band, which enables longer transmission distances and greater power budgets. However, the fundamental limitation of the capacity-distance product of C-band IMDD systems lies in the interplay between fiber dispersion and the square-law detection of photodiodes, which introduces multiple spectral fading notches [5]. To mitigate the power fading, time skew enabled vestigial sideband modulations [6] and phase-diverse receivers [7] have been proposed. Simultaneously, spectral null filling based digital-analog integrating scheme has been developed to enhance the capacity[8].

In this paper, we propose and demonstrate a power fading-aware noise shaping (PFA-NS) technique which enables IMDD systems employing low-resolution digital-to-analog converters (DACs) to achieve expanded capacity. Leveraging the inherent power fading response of C-band IMDD transmission, the proposed PFA-NS incorporates a 2/3-bit digital quantizer with noise feedback modulation to spectrally shape quantization noise and clipping distortion and gathered into the fading-induced spectral notches area and the unused-band. The noise components are inherently suppressed by the inherent frequency-selective fading. Further, numerical evaluations are conducted to validate the effectiveness of the proposed PFA-NS technique. For the IMDD system with 10-km SSMF and 2-/3-bit PFA-NS and DAC, the proposed solution obtains a data rate of 116.4 Gb/s, achieving 94% improvement in data rate compared to conventional systems.

## 2. Principle

To effectively shape quantization noise and clipping distortion and gather within the fading notch areas, the inherent known power fading response of C-band IMDD transmission, denoted as $H(z)$, as shown in the inset of Fig. 1(a), is incorporated into the conventional noise shaping architecture [9] [4], constraining the coefficients of the finite impulse response (FIR) filter, $G(z)$, to values optimized for the fading response. The architecture of the proposed PFA-NS technique is depicted in Fig. 1 (a). Similar channel response dependent noise shaping [4], the principle of PFA-NS is to minimize the difference of transmitted signal $X(z)$ and received signals $Y(z)$. The relationship can be expressed as:

$$Y(z) = X(z) + (1 + G(z))\varepsilon(z) \cdot H(z) \tag{1}$$

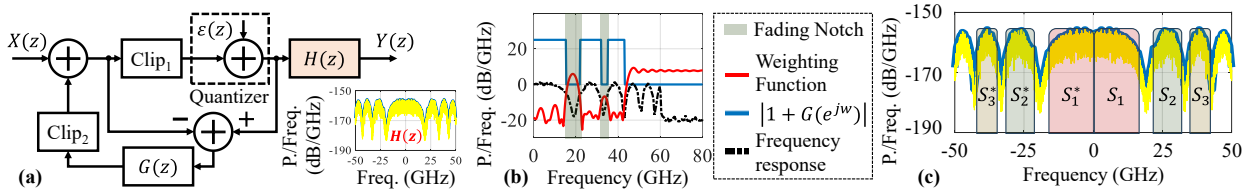

Fig. 1. Architecture of (a) PFA-NS technique. (b) Frequency response and fading notch of C-band IMDD system, and the amplitude response of feedback filter and corresponding weighting function of PFA-NS. (c) Fading-aware frequency division multiplexing.

where $\varepsilon(z)$ represents the quantization noise. $G(z)$ denotes the transfer function of the feedback FIR filter. Thus, the optimal coefficients of $G(z)$ can be determined by minimizing the objective function of [9]

$$\min_G \sum_{i=1}^{N}(1 + G(z)) \cdot H(z) \qquad (2)$$

where $N$ is the number of discretized uniformly spaced frequency points within the signal band. Based on frequency response and fading notches, the amplitude response of the feedback filter, represented by $|1 + G(z)|$, and the corresponding weighting function of the PFA-NS are depicted in Fig. 2. The quantization noise caused by the low-resolution DAC is spectrally shaped and suppressed within the fading-induced notch areas marked by gray shading and the unused spectral regions. The suppressed noise can be filtered by the inherent fading response or the bandwidth limitation of optoelectronic components. Additionally, the $Clip_2$ model is used to prevent noise feedback spikes from saturating the feedforward clipper of $Clip_1$ [9].

Further, in order to expand the capacity, $N$ signals can be fading-aware multiplexed in frequency domain, as shown in Fig. 1(c), followed by aforementioned PFA-NS model.

## 3. Simulation Setup and Results

The simulation setup is illustrated in Fig. 2(a), depicting an external IMDD system with 10 km standard single mode fiber (SSMF) and operating in the C-band, which results in the frequency selective power fading, characterized by the notches shown in Fig. 1(b) indicated by black dot lines. A 2- or 3-bit DAC is employed to reduce hardware cost. According to the specifications mentioned in datasheet of Keysight (M8194A) arbitrary waveform generator, the analog-to digital converter (ADC) is modeled as an ideal quantizer with an effective number of bits (ENOB) of 6 [10]. The electric amplifier (EA) is characterized using the Rapp model of solid-state power amplifier as [11]. To focus on the effects of frequency selective power fading, the back-off value of EA is set to 9 as [12] for the weak nonlinearity. The amplified signal is subsequently loaded into a Mach-Zehnder interference-based intensity modulator (IM) with a half-wave voltage $V_\pi = 4V$. Furthermore, at the receiver side, a variable optical attenuator (VOA) is used to adjust the received optical power (ROP) of target signal. The responsivity, dark current, and the power spectral density of thermal noise of the photodiode (PD) are defined as 0.8 A/W, $5 \times 10^{-9}$A and $10 \times 10^{-12} A/Hz^{1/2}$, respectively.

For the digital signal processing models at the transmitter side (Tx-DSP), as illustrated in Fig. 2(b), based on the frequency response of aforementioned C-band IMDD system, as shown in Fig. 1(b), three independent generated random bit sequences are mapped into the symbols with the format of 4-level pulse amplitude modulation (PAM-4), 16-ary quadrature amplitude modulation (16-QAM) and 16QAM, respectively, as shown in Fig. 2(e). The up-sampling with 2 samples-per-symbol (sps) and root-raised-cosine (RRC) filter with a roll-off factor of 0.1 are used to shape the signals with the symbol rates of 30 Gbaud, 8.1 Gbaud, and 6 Gbaud, respectively. The resampling is performed to maintain a consistent sampling rate of 120 Gsa/s for aforementioned three signals. Subsequently, a power fading-aware frequency-domain multiplexing (FDM) model is used to aggregated the signals for the followed PFA-NS processing. The clipping levels of $Clip_1$ and $Clip_2$ within the PFA-NS model are optimized. The spectrum of the resulting output signal of PFA-NS is presented in Fig. 2(d). The PFA-NS modulated signal is then loaded into a 2- or 3-bit DAC. The amplified signal is externally modulated onto an optical carrier with the central wavelength of 1550 nm, linewidth of 100 kHz and relative intensity noise (RIN) factor of -150 dBc/Hz by the aforementioned IM.

For the DSP models at the receiver side (Rx-DSP), as shown in Fig. 3(c), as the inverse process of aforementioned FDM, power fading-aware frequency domain demultiplexing (De-FDM) is implemented. Additionally, the models of down-sampling to 2-sps, synchronization, $T_s/2$-spaced least mean squares (LMS)-based $2 \times 2$ real-valued equalization, and calculations of the signal-to-noise ratio (SNR) and bit error rate (BER) are used. Furthermore, the soft-decision forward error correction (FEC) scheme with overhead of 37% and 20%, along with a concatenated FEC scheme [13] with overhead of 14.8% are employed, corresponding to the BER thresholds are 2.7E-2, 2.0E-2, and 1.25E-2, respectively.

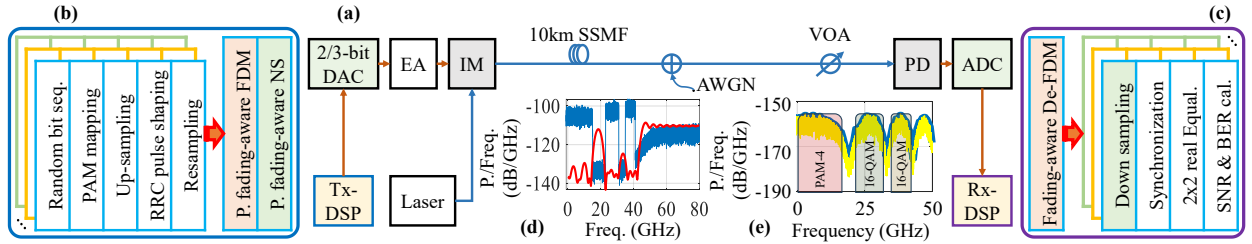

Fig. 2. (a) Simulation setup without optical amplifier and with the (b) Tx-DSP and (c) Rx-DSP. (d) Spectrum of signal from the model of PFA-NS and the weighting function of PFA-NS. (e) Spectrum of power fading-aware FDM signal. EA: electric amplifier; IM: intensity modulator; PD: photodiode

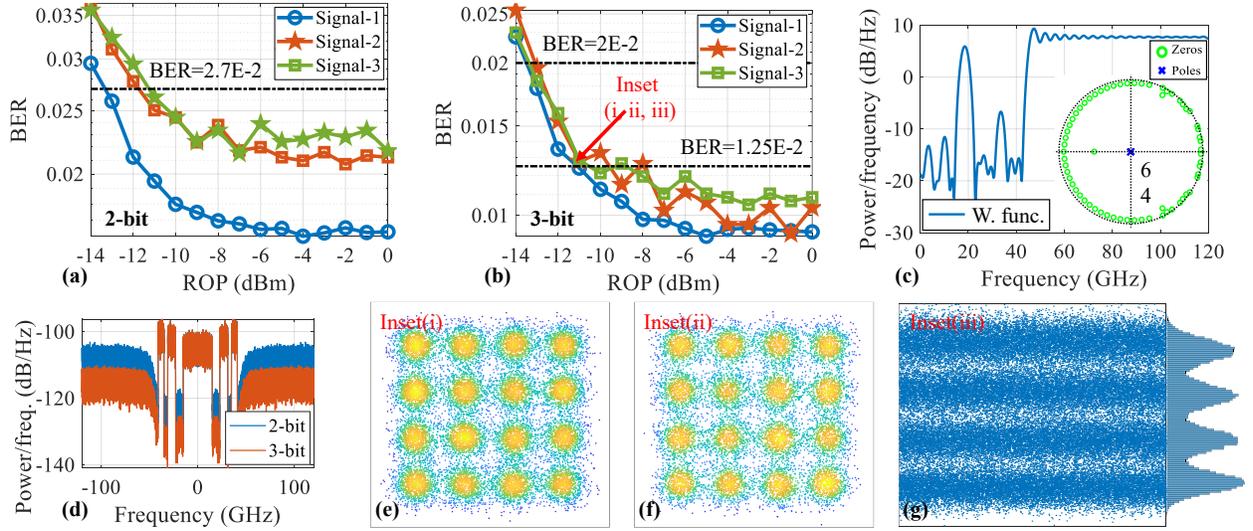

Fig. 3. Measured BER as the function of the received optical power for the (a) 2-bit PFA-NS and DAC, and (b) 3-bit PFA-NS and DAC, respectively. (c) Weighting function and the zeros and poles of PFA-NS. (d) Spectrums of signals from the model of PFA-NS with 2-and 3-bit quantizer. Constellations of the recovered (e) signal-2 and (f) signal-3 for the inset (i) and (ii), respectively. (g) Distribution of the signal-1 for the inset (iii).

In the simulation, the weighting function along with the corresponding zeros and poles of the PFA-NS are shown in Fig. 3(c). The measured BER as a function of received optical power for the 2-bit and 3-bit PFA-NS and DAC are presented in Fig. 3(a) and (b), respectively. Owing to the shaping and concentration of quantization noise within the fading notch areas, the proposed PFA-NS-enabled IMDD system achieves a data rate of 116.4 ($= 30 \times \log_2 4 + 8.1 \times \log_2 16 + 6 \times \log_2 16$) Gb/s at a ROP of -11 dBm. Compared to the conventional IMDD system, which can only transmit signal-1, the proposed solution demonstrates a 94% ($= (116.4 - 60)/60$) improvement in data rate. Compared with Fig. 3(a) against Fig. 3(b), 2-bit quantizer introduces greater quantization noise, degrading the sensitivities of signal-2 and signal-3 with smaller Euclidean distances. Specifically, the spectrums of output signals from the PFA-NS with 2-bit and 3-bit quantizers, along with the constellations and distributions of signal-1, signal-2, and signal-3 are shown in Fig. 3(d), 3(e), 3(f) and 3(g), respectively. Noted that the performance of 2-bit PFA-NS enabled IMDD can be improved by optimizing the power allocation among signal-1, signal-2, and signal-3.

## 4. Conclusions

This paper proposes and demonstrates a PFA-NS technique, which shapes and concentrates quantization noise into the fading notch areas and the unused spectral band, and subsequently filtered by the inherent frequency-selective fading, effectively enabling IMDD systems with low resolution DAC to expand data rate. For an IMDD system utilizing 10 km SSMF with 2-bit or 3-bit PFA-NS and DAC configurations, the proposed solution achieves a data rate of 116.4 Gb/s, obtaining 94% improvement relative to conventional systems.

## 5. Acknowledgements

This work was supported by Doctoral Research Startup Foundation of Xiangtan University under Grant 24QDZ32.

## 6. References

[1] S. Kumar et al., "Intra-data center interconnects, networking, and architectures," in *Optical Fiber Telecommunications VII* (2020).
[2] M. Hermann et al., "2024 ethernet roadmap," IEEE, Koloa, HI, USA, Jan. 2016. Accessed: Sept. 18, 2024.
[3] D. Che et al., "Modulation Format and Digital Signal Processing for IM-DD Optics at Post-200G Era," JLT, 42, 588–605 (2024).
[4] M. Yin et al., "Pre-Equalized DMT Signal Transmission Utilizing Low-Resolution DAC With Channel…," JLT, 41, 3065–3073 (2023).
[5] X. Pang et al., "200 Gbps/Lane IM/DD Technologies for Short Reach Optical Interconnects," JLT, 38, 492–503 (2020).
[6] Y. Zhu et al., "Time skew enabled vestigial sideband modulation for dispersion-tolerant direct–detection …," OL, 45, 6138–6141, (2020).
[7] Y. Zhu et al., "Single-Wavelength 512-Gb/s SSBI-Free Linear Phase-Diverse Direct Detection with Carrier Phase," ICOCN, pp 1-4 (2025).
[8] L. Li et al., "Digital-Analog Hybrid Optical Access Integrating 56-Gbps PAM-4 Signal and 5G mm ….," JLT, 39, 1278–1288 (2021).
[9] W. A. Ling, "Shaping Quantization Noise and Clipping Distortion in Direct-Detection Discrete Multitone," JLT, 32, 1750–1758 (2014).
[10] O. Funke, "M8194A 120 GSa/s arbitrary waveform generator", [Online]. Available: www.keysight.com
[11] V. Bajaj et al., "Efficient Training of Volterra Series-Based Pre-distortion Filter Using Neural Networks," OFC, p. M1H.3 (2022).
[12] X. Zeng et al., "PDA-RoF: Polar Coordinates Assisted Hybrid Digital-Analog Radio-over-Fiber Modulation and …," ICOCN, pp 1-4 (2025).
[13] "Implementation Agreement 400ZR." 2020. [Online]. Available: https://www.oiforum.com